\documentstyle[11pt]{l-aa}

\begin{document}
\voffset=-0.6 in

\title{Rapid fading of optical afterglows as evidence 
       for beaming in gamma-ray bursts}

\author{ Y.F.~Huang\inst{1,2}, Z.G.~Dai\inst{1,3}, and T.~Lu\inst{1,3} }

\offprints{ T.~Lu (E-mail: tlu@nju.edu.cn) }

\institute{ Department of Astronomy, Nanjing University, Nanjing 210093, 
	    P.R. China 
\and 
            Astronomical and Astrophysical Center of East China, 
            Nanjing University, Nanjing 210093, P. R. China
\and  
            IHEP, Chinese Academy of Sciences, Beijing 100039, P. R. China
           }

\thesaurus{13.07.1, 09.10.1, 08.14.1, 02.18.8}

\date{Received  ; accepted }

\maketitle
\markboth{Y.F. Huang et al.: Beaming in GRBs}{}

\begin{abstract}

Based on the refined dynamical model proposed by us earlier for 
beamed $\gamma$-ray burst ejecta, we carry out detailed numerical 
procedure to study those $\gamma$-ray bursts with rapidly fading 
afterglows (i.e., $\sim t^{-2}$). It is found that optical afterglows 
from GRB 970228, 980326, 980519, 990123, 
990510 and 991208 can be satisfactorily 
fitted if the $\gamma$-ray burst ejecta are highly collimated, with 
a universal initial half opening angle $\theta_0 \sim 0.1$. The obvious 
light curve break observed in GRB 990123 is due to the 
relativistic-Newtonian transition of the beamed ejecta, and the rapidly 
fading optical afterglows come from synchrotron emissions during the 
mildly relativistic and non-relativistic phases. We strongly suggest 
that the rapid fading of afterglows currently observed in some 
$\gamma$-ray bursts is evidence for beaming in these cases.

\keywords{Gamma rays: bursts $-$ ISM: jets and outflows $-$ 
          stars: neutron $-$ relativity}

\end{abstract}

\section{Introduction}

The cosmological origin of $\gamma$-ray bursts (GRBs) has been well 
established due to recent discovery of multi-wavelength afterglows 
(Costa et al. 1997; Metzger et al. 1997; Galama et al. 1997; 
Wijers, Rees \& M\'{e}sz\'{a}ros 1997; Piran 1999). However, we are 
still far from resolving the puzzle of GRBs, because their ``inner 
engines'' are well hidden from direct afterglow observations. Some 
GRBs localized by BeppoSAX satellite have implied isotropic energy 
release of more than $10^{54}$ ergs (Kulkarni et al. 1998, 1999; 
Andersen et al. 1999; Harrison et al. 1999), which forces many theorists 
to deduce that GRB radiation must be highly collimated. Obviously, 
whether GRBs are beamed or not has become one of the most important 
problems that need to be solved urgently.

In the literature, it is generally believed that afterglows from jetted 
GRB remnant are characterized by an obvious break in the light curve 
during the {\em relativistic phase}, due to both the jet edge effect 
(Panaitescu \& M\'{e}sz\'{a}ros 1999; Kulkarni et al. 1999; 
M\'{e}sz\'{a}ros \& Rees 1999) and the lateral expansion effect
(Rhoads 1997, 1999). The breaking point is determined by 
$\gamma \sim 1/\theta$, where $\gamma$ is the Lorentz factor of the 
jet and $\theta$ is the half opening angle. Recently we have developed 
a refined dynamical model that can correctly describe the overall 
evolution of an ultra-relativistic jet to non-relativisitic phase with
the expanding velocity 
as small as $10^{-3} c$ (Huang et al. 2000a). Surprisingly enough, our 
detailed numerical results (Huang, Dai \& Lu 2000b) 
show that the break theoretically predicted 
in light curve does not appear during {\em the relativistic phase},
i.e., the time determined by $\gamma \sim 1/\theta$ is not a breaking 
point. However, an obvious break does appear within the 
relativistic-Newtonian transition region, the degree of which is found 
to be parameter dependent (Huang, Dai \& Lu 2000b). Generally speaking, 
the Newtonian phase of jet evolution is characterized by a rapid decay 
of optical afterglows, with the power-law timing index 
$\alpha \sim 1.8$ --- 2.1. 

In practical observations, the power-law decay indices of afterglows 
from GRB 980326, 980519 and 991208 are anomalously large, 
$\alpha \sim 2.0$ (Groot et al. 1998; Owens et al. 1998; 
Castro-Tirado et al. 1999b), and optical light curves of GRB 990123 
and 990510 even show obvious steepening at observing time 
$t \geq 1$ --- 2 d (Kulkarni et al. 1999;  Harrison et al. 1999; 
Castro-Tirado et al. 1999a). Recently GRB 970228 was also reported 
to have a large index of $\alpha \sim 1.73$ (Galama et al. 1999b). 
These phenomena have been widely regarded as evidence for 
beaming (Sari, Piran \& Halpern 1999; Castro-Tirado et al. 1999a). 
The purpose of this {\em Letter} is to study these cases 
numerically, based on our refined beaming model (Huang et al. 2000a).
It is found that optical afterglows from these GRBs can be 
easily reproduced, thus a jet model is strongly favored.

\begin{figure}
\begin{picture}(100,160)
\put(0,0){\includegraphics{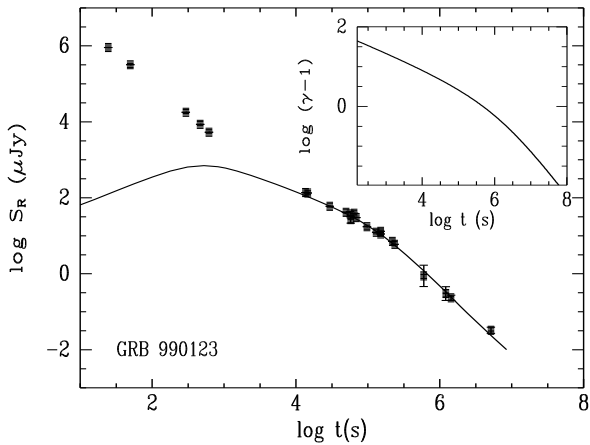}}
\end{picture}
\caption
{ Observed optical (R band) afterglow from GRB 990123 (Galama et al. 
1999a; Kulkarni et al. 1999; Fruchter et al. 1999; GCN 354). The solid 
line is our best fit to it by employing a jet model. Inset shows the 
evolution of $\gamma$ in our model }
\end{figure}

\section{Model}

The importance of non-relativistic expansion phase to our 
understanding of GRB afterglows has been stressed very early by
Huang et al. (1998a, b). A refined generic model that is 
appropriate for both ultra-relativistic and non-relativistic 
isotropic blastwaves has been proposed (Huang, Dai \& Lu 1999a, b). 
Very recently, a similarly refined dynamical model for beamed 
GRB ejecta  was also developed (Huang et al. 2000a). Our 
calculation here will be based on this model. For completeness, 
we describe the model briefly here. For details please see 
Huang et al. (2000a).

Let $R$ be the distance from the burster in the burster frame, 
$M_{\rm ej}$ be the initial ejecta mass, $n$ be the particle number 
density of the surrounding interstellar medium (ISM), and $m$ be 
the swept-up ISM mass. The evolution of the beamed ejecta is 
described by (Huang et al. 2000a):
\begin{equation}
\label{drdt1}
\frac{{\rm d} R}{{\rm d} t} = \beta c \gamma (\gamma + \sqrt{\gamma^2 - 1}),
\end{equation}
\begin{equation}
\label{dmdr2}
\frac{{\rm d} m}{{\rm d} R} = 2 \pi R^2 (1 - \cos \theta) n m_{\rm p},
\end{equation}
\begin{equation}
\label{dthdt3}
\frac{{\rm d} \theta}{{\rm d} t} = 
              \frac{c_{\rm s} (\gamma + \sqrt{\gamma^2 - 1})}{R},
\end{equation}
\begin{equation}
\label{dgdm4}
\frac{{\rm d} \gamma}{{\rm d} m} = - \frac{\gamma^2 - 1}
       {M_{\rm ej} + \epsilon m + 2 ( 1 - \epsilon) \gamma m}, 
\end{equation}
\begin{equation}
\label{cs5}
c_{\rm s}^2 = \hat{\gamma} (\hat{\gamma} - 1) (\gamma - 1) 
	      \frac{1}{1 + \hat{\gamma}(\gamma - 1)} c^2 , 
\end{equation}
where $\beta = \sqrt{\gamma^2 - 1} / \gamma$, 
$m_{\rm p}$ is the proton mass, $c_{\rm s}$ is the co-moving 
sound speed, $\epsilon$ is the radiative efficiency (here we 
consider only adiabatic jets, for which $\epsilon \equiv 0$),
$\hat{\gamma} \approx (4 \gamma + 1)/(3 \gamma)$ is the 
adiabatic index (Dai, Huang \& Lu 1999).

\begin{figure}
\begin{picture}(100,160)
\put(0,0){\includegraphics{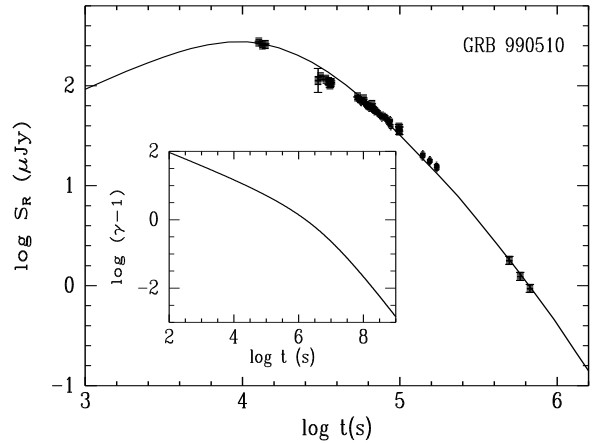}}
\end{picture}
\caption
{ Observed R band light curve of GRB 990510 and our model fit to 
it. Data points were taken from GCN 310, 313, 315, 321, 323, 325,
329 and 332 (also see Stanek et al. 1999; Harrison et al. 1999). Inset 
shows the evolution of $\gamma$ in our model }
\end{figure}

A strong blastwave will be generated due to the interaction of the 
jet and the ISM. Synchrotron radiation from the shock accelerated 
ISM electrons gives birth to afterglows (Sari, Piran \& Narayan 
1998; Vietri 1997a, b). As usual we assume that 
the magnetic energy density in the co-moving frame is a 
fraction $\xi_{\rm B}^2$ of the total thermal energy density 
($B'^2 / 8 \pi = \xi_{\rm B}^2  e'$), and that electrons carry 
a fraction $\xi_{\rm e}$ of the proton energy. This means that the 
minimum Lorentz factor of the random motion of electrons in 
the co-moving frame is 
$\gamma_{\rm e,min} = \xi_{\rm e} (\gamma - 1) 
		     m_{\rm p} (p - 2) / [m_{\rm e} (p - 1)] + 1$, 
where $p$ is the index characterizing the power-law energy distribution 
of electrons, and $m_{\rm e}$ is the electron mass. Our model also 
takes the electron cooling (Sari, Piran \& Narayan 1998) and 
the equal arrival time surface effect (Panaitescu \& 
M\'{e}sz\'{a}ros 1998) into account.

The most important advantage of this model is that it is appropriate 
for both adiabatic and radiative jets, no matter whether their 
expansion is highly relativistic or Newtonian. 

\begin{figure}
\begin{picture}(100,160)
\put(0,0){\includegraphics{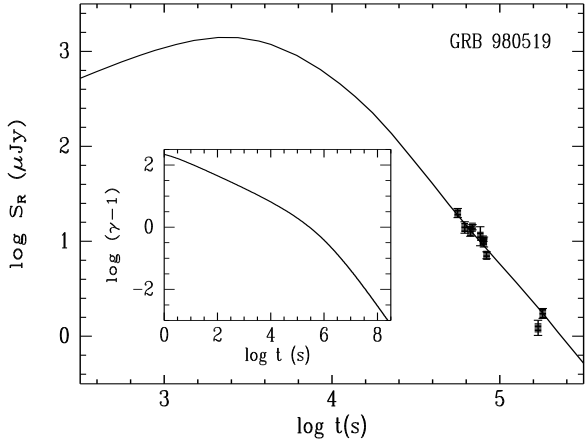}}
\end{picture}
\caption
{ Observed R band light curve of GRB 980519 and our model fit to it.  
Data points were taken from GCN 82, 83, 87, 88, 91, 148, 149 
(also see Halpern et al. 1999). Inset shows the evolution of $\gamma$ 
in our model }
\end{figure}

\section{Numerical Results}

Based on the model described above, 
Huang, Dai \& Lu (2000b) have found that: 
(i) The optical light curve does not break during the relativistic 
phase, i.e., the time determined by $\gamma \sim 1/\theta$ is not 
a breaking point. (ii) An obvious break does appear within the 
relativistic-Newtonian transition region, but its existence depends 
on parameters such as $\xi_{\rm e}, \xi_{\rm B}^2, n, \theta_0$. 
Increase of any of them to a large enough value will make the break 
disappear. (iii) Generally speaking, the Newtonian phase of  the jet 
evolution is characterized by a rapid decay of optical afterglows 
($\alpha \sim 1.8$ --- 2.1). 

In this section we use the model to study those GRBs whose optical 
afterglows decayed rapidly. They include GRB 970228, 980326, 980519, 
990123, 990510 and 991208. We employ a standard Friedmann  cosmology 
with $H_0 = 65$ km s$^{-1}$ Mpc$^{-1}$, $\Omega_0 = 0.2, \Lambda = 0$ 
throughout. 

\subsection{GRB 990123}

GRB 990123 was determined to be at a redshift $z = 1.6004 \pm 0.0008$ 
(Kulkarni et al. 1999), corresponding to a luminosity distance 
$D_{\rm L} \approx 12 $ Gpc. An enormous isotropic $\gamma$-ray 
energy release of $E_{\gamma} \approx 3.4 \times 10^{54}$ ergs was 
estimated (Kulkarni et al. 1999). R band observations made between 
0.16 and 2.75 days after the GRB are well described by a power-law 
with index $\alpha = 1.09 \pm 0.05$ (Fruchter et al. 1999), and the 
light curve steepens greatly when $t \geq 4$ d. Fig. 1 illustrates our
best model fit to the R band light curve. We have taken the following 
initial values and parameters: initial energy per solid angle 
$(E/\Omega)_0 = 3.4 \times 10^{54} {\rm ergs} / 4 \pi$, 
$\gamma_0 = 300$, $n = 10^3$ cm$^{-3}$, $\xi_{\rm e} = 0.2$, 
$\xi_{\rm B}^2 = 10^{-6}$, $\theta_0 =0.13$, $p=2.2$ and $D_{\rm L} = 12$ Gpc. 
In our calculation, the jet enters the sub-relativistic phase at time
$t \sim 10^{5.5}$ s, leading to an obvious break in the light curve. 
We see that observed data points spanning from $t = 0.16$ d to 
$t = 59.5$ d can be well fitted. 

\begin{figure}
\begin{picture}(100,160)
\put(0,0){\includegraphics{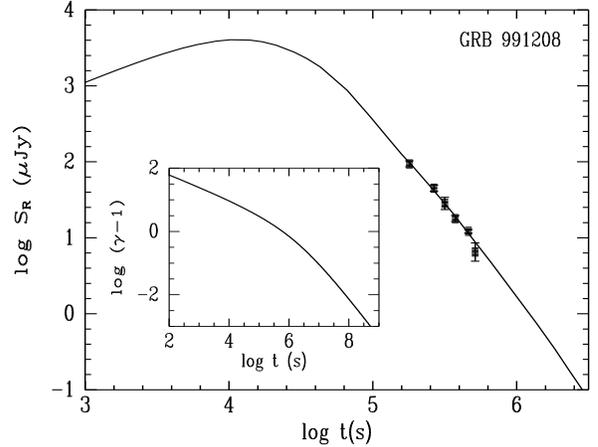}}
\end{picture}
\caption
{ Observed R band light curve of GRB 991208 and our model fit to it. 
Data points were taken from GCN 452, 454, 456, 458, 461 and 
IAU Circ. 7332. Inset shows the evolution of $\gamma$ in our model } 
\end{figure}

\subsection{GRB 990510}

GRB 990510 lies at a redshift $z \geq 1.619 \pm 0.002$ (Vreeswijk et al. 
1999), corresponding to $D_{\rm L} \geq 12$ Gpc. Adopting $D_{\rm L} = 
12$ Gpc implies an isotropic energy of $E_{\gamma} \approx 
2.9 \times 10^{53}$ ergs (Harrison et al. 1999). Fig. 2 illustrates 
R band afterglows reported in the literature, where the solid line is 
our model fit, taking:
$(E/\Omega)_0 = 2.9 \times 10^{53} {\rm ergs} / 4 \pi$, 
$\gamma_0 = 300$, $n = 1$ cm$^{-3}$, $\xi_{\rm e} = 0.2$, 
$\xi_{\rm B}^2 = 0.01$, $\theta_0 =0.075$, $p=2.6$ and $D_{\rm L} = 12$ Gpc. 
In our calculation, the ejecta enters the sub-relativistic phase 
at $t \sim 10^6$ s. But due to relatively large values of $\xi_{\rm e}$ 
and $\xi_{\rm B}^2$, the theoretical light curve peaks very late 
(Huang, Dai \& Lu 2000b), $t_{\rm peak} \sim 10^4$ s, so that observed 
data points before $10^5$ s can be well fitted. For the same reason 
(also see Huang, Dai \& Lu 2000b), 
theoretical flux decays rapidly during the mildly relativistic 
phase (i.e., from $10^5$ s to $10^6$ s), which just meets the 
observational requirements. 

\subsection{GRB 980519}

GRB 980519 had a rapid fading in optical as well as in X-ray band. 
Its optical afterglow is consistent with $t^{-2.05 \pm 0.04}$ 
(Halpern et al. 1999). No redshift was determined. Here we 
assume a typical value of $z \sim 1.4$ for it, corresponding to 
$D_{\rm L} \approx 10$ Gpc. Then the BATSE measured $\gamma$-ray 
fluence of $(2.54 \pm 0.41) \times 10^{-5}$ ergs cm$^{-2}$ implies 
an isotropic energy of $E_{\gamma} \sim 10^{53}$ ergs. Taking
$(E/\Omega)_0 = 1 \times 10^{53} {\rm ergs} / 4 \pi$, 
$\gamma_0 = 300$, $n = 100$ cm$^{-3}$, $\xi_{\rm e} = 0.1$, 
$\xi_{\rm B}^2 = 0.01$, $\theta_0 =0.1$, $p=2.6$ and $D_{\rm L} = 10$ Gpc,
we give our model fit to the R band light curve in Fig. 3. 
In our calculation, due to the relatively large ISM density, the jet 
becomes sub-relativistic at $t \sim 10^5$ s, so the theoretical 
afterglow decays rapidly after $t \sim 10^4$ s. It is clear that 
the rapid fading of GRB 980519 can be reasonably explained in
our model. 

\subsection{GRB 991208}

GRB 991208 was localized by the Interplanetary Network. Beginning 
on 1999 Dec. 10.27 UT, the observed optical afterglow is consistent 
with a rapid decay of $t^{-2.5}$.  A redshift of $z \approx 0.71$ 
was measured (Dodonov et al. 1999), corresponding to 
$D_{\rm L} \approx 4$ Gpc. The implied $\gamma$-ray energy is 
$E_{\gamma} \sim 10^{53}$ ergs. We give our  model fit 
to the R band afterglows in Fig. 4, where we have taken: 
$(E/\Omega)_0 = 1 \times 10^{53} {\rm ergs} / 4 \pi$, 
$\gamma_0 = 300$, $n = 10$ cm$^{-3}$, $\xi_{\rm e} = 0.2$, 
$\xi_{\rm B}^2 = 0.01$, $\theta_0 =0.1$, $p=2.8$ and $D_{\rm L} = 4$ Gpc.
Due to the relatively large values of $n, \xi_{\rm e}$ and $\xi_{\rm B}^2$, 
the theoretical light curve peaks at $\sim 10^4$ s, and it turns into a 
single line when $t \geq 10^5$ s. We see that the observed data points 
can be well fitted. 

\subsection{GRB 970228 and 980326}

GRB 970228 is the first GRB that an optical counterpart was observed. 
Recently, evidence for a supernova was found in the reanalyzed optical 
and near-infrared images. Galama et al. (1999b) argued that the 
afterglow observations are well explained by an initial power-law 
decay with $\alpha = 1.73_{-0.09}^{+0.12}$, modified at later times 
by a type-Ic supernova light curve. Here we suggest that the  initial 
rapid decay of $t^{-1.73}$ (beginning at $t \sim 5.5 \times 10^4$ s) 
can be easily explained if a jet model was employed. This can be 
clearly seen from our calculations in Sections 3.1 --- 3.4. 

GRB 980326 provides the most strong evidence for GRB/Supernova 
connection. Its optical afterglow has two distinct contributions:
a power-law decaying component ($\alpha \sim 2.1$, Groot et al. 1998) 
and emission from the underlying supernova (Bloom et al. 1999). 
Similarly we suggest that the rapidly decaying afterglow component 
(beginning at $t \leq 3.6 \times 10^4$ s) is produced by beamed 
GRB ejecta. 

\section{Discussion and Conclusions}

Determining the existence of beaming in GRBs will be helpful 
to our understanding of the GRB ``central engines''. Based on our refined 
dynamical model for beamed GRB remnants (Huang et al. 2000a), 
we have examined those GRBs with rapidly decaying optical afterglows
closely. Detailed numerical results show that afterglows from 
GRB 970228, 980326, 980519, 990123, 
990510 and 991208 can be satisfactorily 
fitted: the obvious break in the optical light curve of GRB 990123 
is due to the relativistic-Newtonian transition of the beamed ejecta, 
and the rapid fading of afterglows from other GRBs is due to the 
relatively large values of $\xi_{\rm e}, \xi_{\rm B}^2$ and $n$. 
We thus strongly suggest that these GRBs be highly collimated. 
Note that in all cases, synchrotron radiation during the mildly 
relativistic and non-relativistic phases plays an important role in 
explaining the rapid decaying of optical afterglows.

In our calculations, the $\xi_{\rm e}$ values distribute in a 
narrow range, i.e., between 0.1 and 0.2. This has given some 
support to Freedman \& Waxman's (1999) suggestion that $\xi_{\rm e}$ 
has a universal value of $\sim 1/3$. However our results do not 
support their proposal that $p$ has a universal value of 2.2. 
Our $\theta_0$ values distributed in a narrow range, $\theta_0 
\sim 0.1$. This may provide important clues to our understanding 
of the central engine. We note that Woosley et al. (1999) have 
obtained such a small value of $\theta_0$ after numerically 
studying the collapsar models of GRBs. The range of our $n$ values 
($n \sim 1$ --- 1000 cm$^{-3}$) indicates that some GRBs may 
be in gaseous environments, also giving some hints on GRB central 
engines. 

The jet model greatly relaxes the energy crisis for GRB 990123 
and 990510. However, we should keep in mind that other GRBs 
such as GRB 970508, 971214, 980329 and 980703 do not have rapidly 
fading afterglows. They should not be highly collimated (Huang,  
Dai \& Lu 2000b). Then the energy crisis is really a problem: 
GRB 971214 and 980703 have indicated isotropic $\gamma$-ray 
energies of $\sim 0.17$ M$_{\odot} c^2$ and $\sim 0.06$ 
M$_{\odot} c^2$ respectively!

The rapid fading of afterglows from GRB 970228, 980326, 980519 has 
also been explained as being due to the interaction of an isotropic 
blastwave with a wind environment (Dai \& Lu 1998; 
Chevalier \& Li 1999). However 
the wind environment model could not explain the light curve break 
observed for GRB 990123 and 990510. As has been shown clearly in 
this paper, the jet model can naturally explain all these bursts,
and should be more reasonable. 

Another possibility was proposed recently by Dai \& Lu (1999, 2000). 
They suggested that the light curve break is due to the 
relativistic-Newtonian transition of an isotropic blastwave 
in a dense medium ($n \sim 10^6$ cm$^{-3}$), and the rapidly 
fading afterglows can be explained as emissions in the 
non-relativistic phase. It is obviously an interesting proposition 
and should also be paid attention to. 

\acknowledgements 

We thank R. Wijers for his valuable comments. 
This work was partly supported by the National Natural Science 
Foundation of China, grants 19773007 and 19825109, and the National Climbing 
Project on Fundamental Researches.

\end{document}